\documentclass[12pt]{iopart}
\usepackage{graphicx}
\usepackage{bm}
\usepackage{epstopdf}
\usepackage{epsfig}
\usepackage{amssymb,amsmath,stmaryrd,tabularx}
\usepackage[utf8]{inputenc}
\usepackage[center]{subfigure}
\usepackage{amssymb}
\usepackage[unicode=true,pdfusetitle,bookmarks=true,
bookmarksnumbered=false,bookmarksopen=false,breaklinks=false,
pdfborder={0 0 0},backref=false,colorlinks=true,citecolor=red]
{hyperref}
\usepackage{cleveref}
\usepackage[dvipsnames]{xcolor}
\usepackage[normalem]{ulem}

\newcommand{\beq}{\bea}  
\newcommand{\eeq}{\eea}  
\newcommand{\bea}{\begin{eqnarray}}  
\newcommand{\eea}{\end{eqnarray}}  

\newcommand{\rb}{\mathbf{r}}
\newcommand{\ub}{\mathbf{u}}
\newcommand{\vb}{\mathbf{v}}

\begin{document}
\title{The role of fluid flow in the dynamics of active nematic defects}
\author{Luiza Angheluta$^1$, Zhitao Chen$^2$, M. Cristina Marchetti$^2$, Mark J. Bowick$^3$}

\address{$^1$Njord Centre, Department of Physics, University of Oslo, P. O. Box 1048, 0316 Oslo, Norway}
\address{$^2$Department of Physics, University of California Santa Barbara, Santa Barbara, CA 93106, USA}
\address{$^3$Kavli Institute for Theoretical Physics, University of California Santa Barbara, Santa Barbara, CA 93106, USA}

\begin{abstract}
We adapt the Halperin-Mazenko formalism to analyze two-dimensional active nematics coupled to a generic fluid flow. The governing   hydrodynamic equations lead to evolution laws for nematic topological defects and their corresponding density fields.  We find that $\pm 1/2$ defects are propelled by the local fluid flow and by the nematic orientation coupled with the flow shear rate. In the overdamped and compressible limit, we recover the previously obtained active self-propulsion of the  $+1/2$ defects.  Non-local hydrodynamic effects are primarily significant for incompressible flows, for which it is not possible to eliminate the fluid velocity in favor of the local defect polarization alone. For the  case of two defects with opposite charge, the non-local hydrodynamic interaction is mediated by non-reciprocal pressure-gradient forces. Finally, we derive continuum equations for a defect gas coupled to an arbitrary (compressible or incompressible) fluid flow.  
\end{abstract}

\date{\today}

\maketitle

\section{Introduction}
\label{sec:intro}
Nematic order has been widely observed in two-dimensional (2D) realizations of active matter~\cite{marchetti2013hydrodynamics,Doostmohammadi2018}, from vertically vibrated rods~\cite{Narayan2007} to mixtures of cytoskeletal filaments and associated motor proteins~\cite{Sanchez2012,Kumar2018,tan2019topological}, bacterial suspensions~\cite{doostmohammadi2016defect,Nishiguchi2017,copenhagen2020topological}, cell sheets~\cite{Duclos2016,Saw2017,Kawaguchi2017,Blanch2018}, and even developing organisms~\cite{maroudas2020topological}. Nematic order has only quasi-long-range (power-law) falloff in 2D active systems~\cite{chate2006simple,mishra2010dynamic,shankar2018low}, as it does in equilibrium,  and is easily  destroyed by  active stresses that drive spontaneous flows accompanied 
by the proliferation of topological defects~\cite{Sanchez2012}. A distinctive  property of active nematics is that the comet-like $+1/2$ disclination becomes motile~\cite{Sanchez2012,Giomi2013,Pismen2013}, allowing for an activity-driven defect unbinding transition~\cite{shankar2018defect} to a turbulent-like state with chaotic spatio-temporal dynamics. Both experiments and numerical studies based on the solution of continuum models have demonstrated  that this complex dynamics can be characterized by focusing either on the statistics of flow vortices or on that of topological defects~\cite{Doostmohammadi2018,giomi2014defect,thampi2014vorticity,giomi2015geometry,hemingway2016correlation,tan2019topological}.

Recent work has focused on formulating an effective description of defects in active nematics as quasiparticles~\cite{Giomi2013,thampi2013velocity,Pismen2013,giomi2014defect,pismen2017viscous,shankar2018defect,shankar2019hydrodynamics,vafa2020multi}. This approach parallels well-established work in equilibrium systems, where defects can be described as a gas of Coulomb charges and are known to drive order-disorder transitions in 2D~\cite{kosterlitz1973ordering}.

An important new ingredient in 2D active nematics is that the active self-propulsion of the $+1/2$ defects requires taking into account the local geometric polarity of the nematic texture near the defect core~\cite{Giomi2013,Pismen2013,vromans2016orientational,tang2017orientation}. The self-propulsion arises from local active flows generated by the distortion of the texture due to the defect itself, and is absent for the $-1/2$ defects due to their threefold symmetry. Shankar \emph{et al.}~\cite{shankar2018defect} have shown that the dynamics of the gas of unbound defects can be mapped onto that of a mixture of self-propelled (the $+1/2$ defects) and passive (the $-1/2$ defects) quasiparticles with Coulomb interactions and aligning torques, provided that the texture gradients generated at the core of one defect by the other defects can be treated as quasistationary. More recently, the inclusion of multi-defect interactions, within the same quasistatic approximation, has revealed new non-central and non-reciprocal forces in multi-defect textures~\cite{vafa2020multi}. Important corrections to the effective dynamics also arise from the fact that the phase field induced  at the core of a defect by all the others depends on all the defect velocities, as well as on their past history of accelerations, but  including such effects remains a formidable challenge.   These previous works all treat the case of an active nematic on a substrate, where friction dominates over viscous dissipation and the compressible flow~\footnote{Here and below we use the term `compressible flow' to refer to flows that do not satisfy incompressibility (hence $\bm\nabla\cdot\bm{u}\not=0$), although we assume the density to be constant. This could be achieved, for instance, in systems where density is not conserved due to birth and death.} is slaved to the gradient of the  active stress. Recent numerical works \cite{thijssen2020role} show that friction with the substrate promotes the kink wall formation in the trail of $+1/2$ defects and this may have a persistent memory effect on the collective behavior of many defects \cite{nejad2020memory,pearce2020scale}. On the other hand, this limit case of a compressible flow offers an analytically tractable regime, where the flow can be readily eliminated in favor of the nematic texture~\cite{srivastava2016negative,putzig2016instabilities} and  generates active corrections to the nematic stiffness and new nonlinear terms in the equation for the nematic order parameter~\cite{srivastava2016negative,putzig2016instabilities}. In most experimental realizations, however, the flow is incompressible, 
which results in  long-range hydrodynamic effects neglected in previous work. 

In this paper we formulate the dynamics of defects as quasiparticles in the presence of \emph{arbitrary} flows and apply them to flows governed by the Stokes equation, which balances dissipative forces (friction and viscosity) with pressure and active stress gradients. 
We find that, as shown in Ref.~\cite{pismen2017viscous}, incompressibility reduces the active propulsive speed of an isolated $+1/2$ defect. We also show, for the first time, that non-local hydrodynamic effects become important for multi-defect configurations and yield a finite mobility for the $-1/2$ defects. We demonstrate this explicitly by calculating the pressure-gradient forces on a defect dipole and showing that incompressibility gives rise to hydrodynamic interactions that renormalize the propulsion of positive defects and mediate non-reciprocal active defect-defect interactions  additional to the familiar Coulomb force. These couplings due to pressure gradients  are distinct from the perturbative multi-defect interactions obtained in Ref.~\cite{vafa2020multi} for a compressible nematic. 

By coarse graining the defect equations, we also obtain continuum equations for the defect densities and the polarization density of the $+1/2$ defects coupled to an arbitrary fluid flow. We find that the $+1/2$ defects indeed behave like polar particles in a fluid, though advected and rotated by the flow velocity as well as by the Magnus-type force generated by all the other defects~\cite{shankar2019hydrodynamics}. Our work extends the continuum equations obtained in Ref.~\cite{shankar2019hydrodynamics} for a compressible nematic where  flows are slaved to texture deformation. The defect hydrodynamics obtained here  applies to both compressible and incompressible flows and provides a useful starting point for describing experimentally relevant situations with spatially inhomogeneous activity, where defects can be trapped and guided by activity gradients~\cite{shankar2019hydrodynamics,tang2020alignment}, as observed for instance in active actomyosin suspensions~\cite{zhang2019structuring}. 

Our work relies on describing topological defects and their equation of motion as the zeros of an appropriate nematic order parameter. This method  was originally introduced by Halperin~\cite{halperin1981physics} and extended by Mazenko to $O(n)$-symmetric order parameters in the context of phase-ordering kinetics~\cite{mazenko1997vortex}.  It was shown in Refs.~\cite{angheluta2012anisotropic,skaugen2016vortex,skaugen2018dislocation} that this is an efficient numerical tool for tracking the positions and velocities of vortices in superfluids and dislocations in crystals.   The Halperin-Mazenko (HM) method allows us to derive both the discrete defect dynamics and the dynamics of the corresponding defect density fields from the coupled continuum equations for the flow and the nematic order parameter.  For overdamped compressible flows we recover the equations previously derived in~\cite{shankar2018defect}. 

The rest of the paper is structured as follows. In Sec.~\ref{sec:model} we briefly summarize the well-established continuum model of active nematic hydrodynamics. In Sec. \ref{Sec:HM}, we present the HM method based on the conservation of topological defects as zeros of the order parameter and relate the defect velocity to the nematic texture. Here we also  derive the equation of motion for the polarization of a $+1/2$ defect in the quasistatic-phase approximation. In Sec. \ref{sec:flows}, we evaluate the defects propulsive velocity arising from activity-induced flows. We show that our method recovers previous results for isolated defects for both compressible~\cite{Giomi2013,Pismen2013,shankar2018defect} and incompressible~\cite{pismen2017viscous} flows. We also present new results for a defect dipole in an incompressible nematic, where flows driven by pressure gradients yield new contributions to the motility of both the positive and the negative defect, as well as non-reciprocal active forces.  In Sec. \ref{Sec:FieldEqs} we derive general hydrodynamic equations for a defect gas, applicable to both compressible and incompressible flows. Finally, we summarize and conclude in Sec. \ref{Sec:conclusion}. The Appendices detail analytical computations of the flow field induced by a variety of specific defect configurations. 

\section{Continuum model of 2D active nematics}
\label{sec:model}
In continuum models, the hydrodynamics of uniaxial active nematics is described by the $Q$-tensor $\textendash$ the order parameter for the orientational field $\theta(\rb,t)$ and its associated nematic director (line field) $\hat n = [\cos\theta, \sin\theta]$. The head-tail symmetry of the local director ($\hat n \equiv -\hat n$) and the property that $Q$ vanishes in the isotropic phase constrain the $Q$-tensor to be symmetric and traceless. In 2D, $Q_{ij} = S(2\hat n_i \hat n_j-\delta_{ij})$ with $S$ the magnitude of the nematic order. $Q$ has two independent components $Q_{xx}=-Q_{yy}$ and $Q_{xy} = Q_{yx}$. The evolution of the $Q$-tensor is given by the the Edwards-Beris equation, which in 2D reads 
\cite{genkin2017topological,marchetti2013hydrodynamics}
\begin{flalign}\label{eq:Q_active}
&(\partial_t +\bm u\cdot\nabla) Q_{ij} = \lambda E_{ij} +\lambda'Q_{ij}\partial_ku_k+ Q_{ik}\Omega_{kj}-\Omega_{ik} Q_{kj}+ \frac{1}{\gamma}\left[K \nabla^2 Q_{ij}+g(1-Q_{kk}^2)Q_{ij}\right].
\end{flalign}
The last two terms on the right hand side (RHS) represent the relaxation of the order parameter to minimize isotropic distortions, with elastic constant $K>0$, and  deviations from the uniform ordered state, $S_0^2 =1$. Here $\gamma$ is the rotational friction coefficient of the nematic. The remaining terms are given by the material time derivative in the presence of the fluid flow velocity $\ub$ with vorticity $2\Omega_{ij} = \partial_i u_j-\partial_j u_i$ and the tendency of nematics to align with the flow through linear coupling to the symmetric traceless part of the flow strain rate, $2E_{ij} = \partial_i u_j+\partial_j u_i-\delta_{ij}\partial_k u_k$, and to the flow divergence, $\bm\nabla\cdot\bm{u}$. We have neglected higher-order flow alignment terms which are nonlinear in the order parameter. The flow alignment parameters $\lambda$ and $\lambda'$ are dimensionless but nonuniversal. They are controlled by the degree of nematic order and the microscopic shape of the underlying mesogens. The alignment with shear flow, controlled by $\lambda$, directly affects the defect velocity, whereas the coupling to the flow divergence, controlled by $\lambda'$, is linear in the $Q$-tensor, which vanishes at the defect core. As a result this term may be neglected in determining quasistatic defect dynamics.

Flows in active nematics are generally characterized by very low Reynolds numbers and follow the Stokes equation, 
\beq\label{eq:u_stokes_unrescaled}
\mu \bm u= \eta_0\nabla^2\ub-\bm\nabla \Pi +\bm\nabla\cdot \bm\sigma^a\;,
\eeq
describing force balance between friction with the substrate ($\mu$), viscous dissipation controlled by the shear viscosity $\eta_0$,  potential forces due to pressure, $\Pi$, and active stress, $\bm\sigma^a$. Pressure becomes important for incompressible flows, when it is determined by the incompressibility constraint $\bm\nabla\cdot \bm u=0$. The active stress is proportional to the order parameter $\bm\sigma^a = \alpha_0 \bm{Q}$, where $\alpha_0$ is the activity coefficient, such that $\alpha_0<0$ corresponds to extensile activity, and $\alpha_0>0$ to contractile activity. Finally, in Eq.~\eqref{eq:u_stokes_unrescaled} we neglect elastic liquid crystalline stresses that are higher order in the gradients of the order parameter field compared to the active stress.

Dimensional analysis reduces the parameter space to two independent dimensionless numbers. We rescale time $t = \tau \tilde{t}$ by the nematic relaxation time $\tau = \gamma/g$, and space $\rb = \xi \tilde{\rb}$ by the nematic coherence length $\xi = \sqrt{K/g}$, the intrinsic length scale that determines the defect core size and the minimal length scale of spatial variations of the order parameter. The fluid velocity then scales as $\ub = (\xi/\tau) \tilde \ub$ and the fluid pressure as $\tilde{p}=p/(\mu \xi^2/\tau)$. The relevant parameters for the dimensionless equations then become the rescaled activity $\alpha = \alpha_0/(\mu \xi^2/\tau)$, measuring the strength of the active force relative to friction, and the rescaled shear viscosity $\eta=\eta_0/ (\mu\xi^2)$, measuring viscous forces relative to friction. From now on we will drop the tildes on the dimensionless variables and treat the friction-dominated regime $\eta\rightarrow 0$. The general case with an interplay between friction and viscous dissipation will be treated separately elsewhere.

The nematic order field has an elegant complex representation  $\psi = S e^{i2\theta} = Q_{xx}+iQ_{xy}$ \cite{oza2016antipolar,pismen2017viscous,shankar2019hydrodynamics}.  The dynamics of the complex $\psi$-field following 
from Eq.~\eqref{eq:Q_active} is given by a generalized complex Ginzburg-Landau (CGL) equation 
\begin{flalign}\label{eq:psi_complex}
\partial_t \psi +u\partial_z \psi+\bar u\partial_{\bar z} \psi  = \lambda \partial_{\bar z} u +(\partial_z u-\partial_{\bar z} \bar u)\psi+4 \partial_z\partial_{\bar z}\psi + (1-|\psi|^2)\psi,
\end{flalign}
with  
\beq\label{eq:u_stokes}
u =  4\eta\partial_z\partial_{\bar z}u+2 \alpha \partial_z\psi - 2\partial_{\bar z} \Pi,
\eeq
and the incompressibility condition  $\partial_{\bar{z}}u+\partial_z\bar{u}=0$, with $z= x+iy$ and $\partial_z = (\partial_x-i\partial_y)/2$.

Topological defects appear in the $\psi$-field as localized regions (of size $\xi$) where  $S$ vanishes and around which the nematic orientation $\theta$ winds by $\pi$. The method described in the next section allows us to derive the kinematics of the $\pm 1/2$ nematic defects from the dynamics of the zeros of the $\psi$-field. 

\section{Defect kinematics}
\label{Sec:HM}

The complex nematic order parameter $\psi$ is always a single-valued function even in the presence of defects. This restricts any phase discontinuity in $\theta(\rb)$ to have a half-integer fundamental winding number.  Taking an arbitrary contour $C$ surrounding a set of elementary defects, the net phase jump is  
\bea\label{eq:charge}
\oint_C d\theta  = \oint_C \nabla \theta \cdot d\bm r = 2\pi \sum_{\beta\in C} q_\beta \;,
\eea
where the sum runs over all defects $\beta$ contained inside $C$. We consider only energetically stable defects of charge $q_\beta = \pm 1/2$. Higher charge defects are topologically stable in 2D but are energetically unstable to disassociation into elementary defects, since their intrinsic energy grows as the square of the charge~\cite{chandrasekhar1986structure}. 

Defects are accurately located by the zeros of the complex order parameter. Their dynamics is typically derived by solving the hydrodynamic field equations in the comoving frame of the defect together with asymptotic matching of inner- and outer-core solutions under appropriate initial and boundary conditions \cite{neu1990vortices,pismen1990mobility,Pismen2013}. The defect velocity and the phase gradients are then related through the mobility coefficient which acquires a logarithmic dependence on velocity. The field solutions are derived perturbatively in phase gradients and defect velocity, and are challenging to generalize when a generic fluid flow is coupled with the evolution of the order parameter. The HM method \cite{mazenko1997vortex, liu1992defect},  in which the defect dynamics is identified with the motion of the zeros of the complex order parameter, allows us to directly include the coupling to fluid flow. 

The starting point is to represent a configuration of defects by the Dirac delta function $\delta(\psi)$, which picks up the defect positions $\rb_{\beta}$, with $\beta = 1,2,\cdots$ the particle number index. The field variables map into discrete defect positions $\{\rb_\beta\}$, 
\bea
\sum_\beta \delta (\bm r-\bm r_{\beta}(t)) = |D| \delta[\psi(\rb,t)]\;,
\eea
where the Jacobi determinant $D(\bm r,t) = \det(\partial \psi)$ is 
\bea\label{eq:D_real}
D(\bm r,t)  = \frac{1}{2i} \epsilon_{ij}\partial_i\bar\psi\partial_j\psi \;,
\eea
with $\epsilon_{ij}$   the Levi-Civita anti-symmetric tensor. In complex coordinates the determinant is   
\bea\label{eq:D_complex}
D (z,t) =  \partial_z\psi \partial_{\bar z} \bar \psi-\partial_z \bar\psi \partial_{\bar z} \psi\;.
\eea
The field $D$ captures the zeros of the order parameter and vanishes everywhere except at the defect positions. It obeys a conservation law determined by the hydrodynamic equation of the $\psi$-field and readily obtained by taking time derivatives on both sides of Eq.~\eqref{eq:D_real} \cite{mazenko1997vortex}, with the result
\beq\label{eq:D_conserv}
\partial_t D (\rb,t) + \partial_i J_i^{(\dot\psi)}=0 \;, 
\eeq
where the conservative current $J_i^{(\dot\psi)}$ is given by
\bea\label{eq:J_D}
J_i^{(\dot \psi)} (r,t)= \epsilon_{ij} \, \textrm{Im}(\partial _t\psi\partial_j\bar\psi)\;,
\eea
or in  complex form  
\bea\label{eq:J_dotphi_complex}
J^{(\dot\psi)} &= &J_x^{(\dot\psi)}+iJ_y^{(\dot\psi)} = -\partial_{\bar z} \bar \psi \partial_t \psi+\partial_{\bar z}\psi \partial_t \bar\psi\;.
\eea

The defect charge is related to the sign of the determinant evaluated at the defect position, $q_\beta =\frac{1}{2}\, \text{sgn}(D)|_{\bm r=\bm r_\beta}$\cite{liu1992defect}. This relation allows us to connect the determinant field $D$ with the defect charge density field for a given configuration of defects, namely 
\bea\label{eq:rho}
\rho(\rb,t) = \sum_\beta q_\beta \delta (\rb-\rb_{\beta}(t)) = \frac{1}{2} D(\rb,t) \delta[\psi(\rb,t)]\;.
\eea
The defect charge density $\rho$ also satisfies a conservation law  \cite{mazenko1997vortex,mazenko2001defect}: 
\begin{eqnarray}
\partial_t \rho + \partial_i J_i^{(\rho)} =0\;,
\end{eqnarray}
where the current density is
\bea\label{eq:J_rho}
\bm{J}^{(\rho)} (\bm r,t)= \frac{1}{2}\bm J^{(\dot\psi)} (\bm r,t)\delta(\psi) = \sum_\beta \frac{\bm J^{(\dot\psi)}(\bm r_{\beta})}{D(\bm r_{\beta})} q_\beta \delta(\bm r-\bm r_{\beta})\;. 
\eea

Finally, for a given defect configuration, the density current can be also expressed in terms of the velocity of the defects $\bm v_{\beta} = \dot \rb_{\beta}$:
\beq
\bm J^{(\rho)} (\rb, t)= \sum_\beta q_\beta \bm v_\beta \delta(\rb-\rb_{\beta})\;.  
\eeq
The defect velocity is therefore determined by the field $D$ and its current density:
\bea\label{eq:defect_motion}
\bm v_{\beta} (t) = \frac{\bm J^{(\dot\psi)}}{D}\Big|_{\rb=\rb_\beta} \;,
\eea
which explicitly relates the defect velocity to derivatives of $\psi$ evaluated at defect positions. In complex form $v_\beta = v_{x,\beta}+iv_{y,\beta}$ is given by
\begin{eqnarray}\label{eq:dot_r_complex}
v_\beta &=&   \frac{-\partial_{\bar z} \bar \psi \partial_t \psi+\partial_{\bar z}\psi \partial_t \bar\psi}{\partial_z\psi \partial_{\bar z} \bar \psi-\partial_z \bar\psi \partial_{\bar z} \psi}\Big|_{z=z_\beta}\;.
\end{eqnarray}
This is the starting point for obtaining an explicit form for the defect velocity in terms of the flow field and nematic distortions, using the profile of  $\psi$ in the vicinity of the defect. 

\subsection{Defect velocity}
To relate  the defect velocity to nematic distortions and to flow, we need to evaluate gradients of $\psi$ at the defect core (Eq.~\eqref{eq:dot_r_complex}). Focusing on the $\beta$-th defect, the order parameter can be written as 
\beq
\psi(z,\bar z) = \psi_0(z,\bar z)~e^{2i\tilde{\theta}(z,\bar z)}\;,
\label{eq:psi}
\eeq
where 
\beq
\psi_0(z,\bar z) =S(|z-z_\beta|) \left(\frac{z-z_\beta}{\bar z-\bar z_\beta}\right)^{q_\beta}
\eeq
and $\tilde \theta$ is the smooth phase distortion due to all other defects and to boundary conditions.
The core function $S(|z|)$ has the generic asymptotic behaviors: $S(|z|)\approx |z|$ for $|z|\ll 1$ and  $S(|z|)\approx 1$ for $|z|\gg 1$. Notice that in Eq.~\eqref{eq:dot_r_complex}, the density current and $D$-field, which depend on the derivatives of  $\psi$, are evaluated at the defect position where $S(|z|=0)=0$. To leading order in the phase gradients it is then sufficient to consider only the linear profile of the core function near the defect. Including the full-nonlinear profile will introduce additional contributions from the core structure. 

From Eq.~\eqref{eq:D_complex}, the determinant $D$ at the defect position is simply  $D|_{z=z_\beta} = 2q_\beta$. The defect velocity from Eq.~\eqref{eq:dot_r_complex} is determined directly from the evolution of the $\psi$-field. At the defect core, we have $\partial_t \psi |_{z=z_\beta} = [4\partial_z\partial_{\bar z} \psi-u\partial_z\psi-\bar u\partial_{\bar z}\psi +\lambda \partial_{\bar z} u]_{z=z_\beta}$, since all the local terms in $\psi$ vanish. As a result, the defect velocity does not depend on the local fluid vorticity and ordering potential, and  is given by 
\begin{eqnarray} \label{eq:v_general}
v_{\beta}^{+} &=&  \left[- 8i \partial_{\bar z} \tilde\theta +u -\lambda e^{-2i\tilde\theta}\partial_{\bar z} u\right]_{z=z_{\beta}^{+}}\;,\nonumber\\
v_{\beta}^{-} &=&  \left[+8i \partial_{\bar z} \tilde\theta +u -\lambda e^{2i\tilde\theta}\partial_{z} \bar u \right]_{z=z_{\beta}^{-}}\;.
\end{eqnarray}
The nematic phase $\tilde\theta_\beta= \tilde\theta(z_\beta(t),\bar z_\beta(t)|\{z_\gamma(t),\bar z_\gamma(t)\})$, where $\gamma\not=\beta$ labels all the other defects,  is evaluated at the position of the tagged defect. 
To zeroth-order in activity $\alpha$, the first term on the RHS of each of Eqs.~\eqref{eq:v_general} represents the Coulomb interactions with the other defects. This is easily seen by noting that for $\alpha=0$ the steady-state nematic phase field can be written as the superposition of the phase induced by each point defect~\cite{vafa2020multi},
\begin{eqnarray}
\tilde\theta_{\beta} = \frac{i}{2} \sum_{\gamma \neq \beta} q_\gamma \log\left(\frac{\bar z_{\beta}-\bar z_\gamma}{z_{\beta} -z_\gamma} \right)+\theta_0\;,
\label{eq:thetabeta}
\end{eqnarray}
where $\theta_0$ is a fixed external phase.  In particular, when the system is topologically neutral, $\theta_0$ is equal to the far field orientation of the nematic texture. The first term on the RHS of Eq.~\eqref{eq:thetabeta} determines the Coulomb pairwise interaction force $F_{\beta,\gamma}$ between any two defects  $\beta$ and $\gamma$ as
\beq
 -16 i q_\beta \partial_{\bar{z}_{\beta}}\tilde\theta_\beta =\sum_{\gamma\not=\beta} \frac{8 q_\beta q_\gamma}{\bar z_\beta-\bar z_\gamma}\equiv \sum_{\gamma\not=\beta} F_{\beta,\gamma}\;.
 \label{eq:Coulomb}
\eeq
 The other two terms on the RHS of each of Eqs.~\eqref{eq:v_general} represent the contribution to the defect velocity from the flow velocity $u$.  The terms linear in $u$ describes advection of  both positive and negative  by flows at defect cores. The terms proportional to flow gradients arise from the shear flow alignment, and drive  defect motion in a direction determined by the flow gradient field relative to the local nematic orientation. 

\subsection{Defect polarity}
It has been shown  that defect polarity plays an important role in the dynamics of defects in active nematics~ \cite{shankar2018defect,shankar2019hydrodynamics,vafa2020multi}. The polarity $\bm{e}_{\beta}^+$ of the $\beta$-th $+1/2$ defect is determined by gradients of the order parameter  
\bea
\bm{e}_{\beta}^+=\left[\bm\nabla\cdot\bm{Q}\right]_{\bm{r}=\bm{r}_{\beta}^+}~~~~~~{\rm or}~~~~~~ e_{\beta}^+ = 2[\partial_z\psi]_{z=z_{\beta}^+} = 2 e^{2i\tilde\theta_{\beta}^{+}}\;.
\label{eq:polarity+}
\eea
For a $-1/2$ defect, $\bm\nabla\cdot\bm{Q}$ vanishes at the defect core due the defect's trifold symmetry. The polarity of a $-1/2$ defect can, however, be defined as a triatic~\cite{tang2017orientation} or as a vector $\mathbf{e}_\beta^-$ that points along one of the three equivalent directions that define the defect's orientation~\cite{vromans2016orientational}, defined  as
\bea
&&\left(\mathbf{e}_\beta^- \right)^3=[\bm\nabla \cdot (\bm \sigma_3 : \mathbf{Q})]_{\bm{r}=\bm{r}_{\beta}^-} ~~{\rm or}~~e_{\beta}^- = [2(\partial_{\bar z}\psi )_{z=z_{\beta^-}} ]^{1/3} =\left( 2e^{2i\tilde\theta_\beta^-}\right)^{1/3}\;,
\label{eq:polarity-}
\eea
where $\bm\sigma_3$ is the Pauli matrix, $\bm\sigma_3 = \begin{pmatrix} 1 &  0\\  0 & -1\end{pmatrix}$.

\begin{figure}[t]
\includegraphics[width=0.9\textwidth]{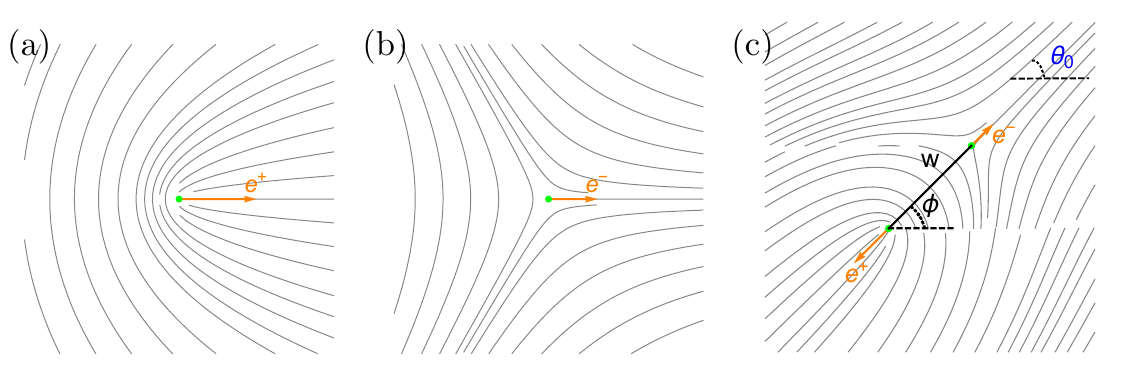}
\caption{Defect polarities $e^{\pm}$ as defined in Eqs.~\eqref{eq:polarity+} and~\eqref{eq:polarity-}. (a) Isolated $+1/2$ defect. (b) Isolated $-1/2$ defect. (c) Dipole with separation $w$. Notice that single defects give rise to long-range distortions in the nematic orientation, whereas the dipole induces short-ranged distortions such that the nematic orientation is uniform in the far-field and given by $\theta_0$.}
\label{fig:polarization}
\end{figure}

In general the polarity of a defect depends on the orientation and position of all other defects, as well as on  $\theta_0$.  Near a  $+1/2$ defect located at the origin and far from all other defects, the order parameter takes the form  $\psi_+(z, \bar z)=(z\bar z)^{1/2}\left(\frac{z}{ \bar z}\right)^{1/2}e^{2i\theta_0}$ and the polarization $e^+=2e^{2i\theta_0}$ is determined entirely by the boundary conditions, as shown in Fig.~\ref{fig:polarization}(a). Similarly, near an isolated $-1/2$ defect located at the origin, $\psi_-(z, \bar z)=(z\bar z)^{1/2}\left(\frac{z}{ \bar z}\right)^{-1/2}e^{2i\theta_0}$ and the associated polarization is $e^-=2^{1/3} e^{2i\theta_0/3}$ (see Fig.~\ref{fig:polarization}(b)). Whereas, for a  dipole consisting of a positive defect at $z^+$ and a negative defect at $z^-$, we have 
$\psi_\pm(z,\bar z)=S(|z-z^+|)S(|z-z^-|)\left(\frac{z-z^+}{ \bar z-\bar z^+}\right)^{1/2}\left(\frac{z-z^-}{ \bar z-\bar z^-}\right)^{-1/2} e^{2i\theta_0}$. Denoting by $w=z^- - z^+ = |w|e^{i\phi}$ the separation between the two defects, the respective polarizations depend both on the far-field boundary condition and the dipole orientation as given by
$e^+=2e^{i[-\phi+\pi+2\theta_0]}$  and $e^-=2^{1/3}e^{i[\phi+2\theta_0]/3}$, as shown in Fig.~\ref{fig:polarization}(c).

We evaluate the dynamics of a given defect assuming that $\tilde\theta(z, \bar z; \{z_\gamma,\bar z_\gamma\})$ is quasistatic, meaning that it is constant in time during the motion of the tagged defect or of any of the other defects. The defect polarization is then not an independent variable since its dynamics is determined by the dynamics of $\{z_\beta(t)\}$. Nonetheless, it provides a useful effective degree of freedom for describing the dynamics of defects as quasiparticles, as well as the hydrodynamics of the defect gas~\cite{shankar2018defect,shankar2019hydrodynamics}.

For a positive  defect, the time evolution of the defect polarity follows simply from taking the time derivative of Eq.~\eqref{eq:polarity+}, with the result~\cite{vafa2020multi}
\begin{eqnarray}\label{eq:edot}
\dot e_\beta^+ &=& \frac{i}{2} e_\beta^+ \textrm{Im} \sum_{\gamma \neq \beta} (v_\beta^+ - v_\gamma) \bar F_{\beta,\gamma} \;,  
\end{eqnarray}
where $v_\beta^+$ is the defect velocity and $F_{\beta,\gamma}$ the pair Coulomb force defined in Eq.~\eqref{eq:Coulomb}. 

\begin{figure}[t]
\includegraphics[width=0.8\textwidth]{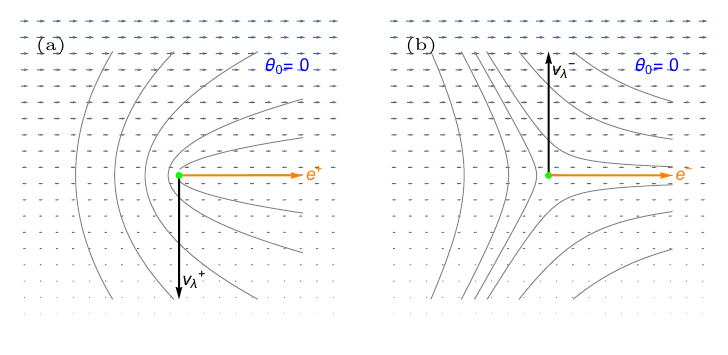}
\caption{The contributions $v_\lambda^{+}=-\frac{\lambda}{2} \bm{E}\cdot \bm{e}_{\beta}^{+}$ and $v_\lambda^{-}=-\frac{\lambda}{2} \bm{E}^\sigma\cdot (\bm{e}_{\beta}^{-})^3$ (black arrows, see Eq.~\eqref{eq:velocity_eqs_gen}) to the velocity of defects induced by shear flow  is shown for specific configuration for a $+1/2$ (panel (a) ) and a $-1/2$ defect (panel (b) ). The orange arrows are the defect polarizations as defined in Eqs.~\eqref{eq:polarity+} and~\eqref{eq:polarity-}. The background arrows represent the flow velocity gradient.}
\label{fig:alignment}
\end{figure}

It is also useful to rewrite the equations governing the defect  dynamics  in real coordinates. Namely, the defect velocity from Eq. \eqref{eq:v_general} in real coordinates reads as 
\begin{eqnarray}  \label{eq:velocity_eqs_gen}
 \bm v_\beta^{+} &=&  \left[4\bm\epsilon\cdot\bm\nabla\tilde\theta + \bm u -\frac{\lambda}{2} \bm{E}\cdot \bm{e}_{\beta}^{+} \right]_{\bm r=\bm r_\beta^{+}}\;, \nonumber\\
\bm v_\beta^{-} &=&  \left[-4\bm\epsilon\cdot\bm\nabla \tilde\theta + \bm u - \frac{\lambda}{2}   \bm E^\sigma \cdot\left(\mathbf{e}_\beta^-\right)^3 \right]_{\bm r=\bm r_\beta^{-}}\;,
\end{eqnarray}
where $\bm E^\sigma = \bm\sigma_3 : \bm E$. The coupling to shear flow drives the $\pm 1/2$ defects to move in a direction controlled by the angle between their polarity and the flow gradient. For example, when the defect polarization points either along the direction of flow or along the direction of flow gradient, the coupling to shear flow drives the $\pm 1/2$  defects to move normal to their polarization, and in opposite directions, as shown in Fig.~\ref{fig:alignment}.

Similarly the equation for the polarization of the $q_\beta = +1/2$ defect takes the real space form
\begin{eqnarray}
\dot{\bm e}_{\beta}^{+} &=&  -\bm\epsilon\cdot\bm e_{\beta}^{+} \frac{1}{2}\sum_{\gamma \neq \beta}\bm F_{\beta,\gamma}\cdot \bm\epsilon \cdot (\bm v_\beta^{+} -\bm v_\gamma)\;. 
\label{eq:torque-real}
\end{eqnarray}
The first term on the RHS of Eq.~\eqref{eq:torque-real} shows that the  polarity of the $+1/2$ defect rotates at a rate dependent on the degree of alignment between defect velocity and the nematic distortion, as found in ~\cite{shankar2018defect}. Equivalently,  the net torque is given by the cross-product of the Coulomb force with the relative velocity of the tagged positive defect with respect to the rest of defects. A result of this form was recently obtained in Ref.~\cite{vafa2020multi} for the case of compressible flow. On the other hand, Eq.~\eqref{eq:torque-real}  holds for any flow condition since flow enters implicitly through the defect velocity. In the next section we obtain a set of closed equations for the defect positions and orientation by expressing the Stokes flow in terms of the defect degrees of freedom. 

\section{Defect-induced flow}
\label{sec:flows}
In this section we obtain explicit expressions for the flow induced by specific defect configurations, which, in turn, drives defect dynamics via Eq.~\eqref{eq:velocity_eqs_gen}. By eliminating the flow in favor of defect degrees of freedom we can identify the advection at the defect core as a defect propulsion velocity. We consider both compressible and incompressible ($\nabla\cdot \ub = 0$) flows. In the latter case, we calculate for the first time  the flow induced by a defect dipole, which was previously studied numerically in Ref.~\cite{Giomi2013}.

\subsection{Single defect in compressible flow}
For compressible flows given only by the active stress $u = 2\alpha \partial_z\psi$, the flow velocity at a positive defect is simply proportional to the defect's polarization $e_{\beta}^+$, while it vanishes at a negative defect. The flow alignment term in Eq.~\eqref{eq:velocity_eqs_gen} can be expressed in terms of the rotated phase gradients using Eq. \eqref{eq:psi} for the profile of $\psi$ near the defect core. This yields the  defect propulsion velocity obtained in previous work \cite{Giomi2013,giomi2014defect,Pismen2013,shankar2018defect,shankar2019hydrodynamics,vafa2020multi}, with defect dynamics governed by 
\begin{eqnarray}  \label{eq:v_comp}
\bm v_{\beta} &=&  8 q_\beta K_\lambda \left(\bm\epsilon\cdot\bm\nabla \tilde\theta \right)_{\bm r=\bm r_{\beta}} + \alpha \bm e_{\beta} \delta_{\beta, \frac{1}{2}}\;,
\end{eqnarray}
where $K_\lambda = 1+\lambda\alpha/2$ is the effective elastic constant (in dimensionless units) incorporating the effect of  flow alignment. 

The  dynamics of the polarization can be obtained from Eqs.~\eqref{eq:torque-real} and ~\eqref{eq:v_comp} as
\begin{eqnarray}\label{eq:edot_0}
\dot{\bm e}_{\beta}^{+} &=&  -\bm\epsilon\cdot\bm e_{\beta}^{+} \left[\frac{\alpha}{2K_\lambda} \bm e_{\beta}^{+}\cdot\bm\epsilon\cdot\bm v_{\beta}^{+} \right] 
\end{eqnarray}

and has precisely the form of the active torque given in Ref.~\cite{shankar2018defect}, but with $K$ replaced by $K_\lambda$. Note that $K_\lambda$ can change sign for extensile activity ($\alpha<0$), resulting in a change of sign of the active torque. Such  corrections are, however, nonperturbative in activity.

\subsection{Incompressible Flow}
For incompressible flows, evaluating the flow velocity requires solving the Poisson equation for the pressure:  
\bea
\partial_z\partial_{\bar z} \Pi = \textrm{Re} (\partial_z^2\psi)\;,  
\label{eq:Poisson-eq}
\eea
where the source term  acquires contributions from the topological defects and any  smooth deformations of the order parameter, such as kink walls.  The solution of Eq.~\eqref{eq:Poisson-eq} can be written as the convolution of the logarithmic Green's function $G(\bm r) = 1/(2\pi) \log(|z|)$ with the source term: 
\bea
\Pi  = 4 \alpha \int dz' d\bar z' G(|z-z'|) \textrm{Re} (\partial_{z'}^2\psi)\;.
\eea
The corresponding incompressible velocity field follows from Eq.~\eqref{eq:u_stokes}:
\begin{eqnarray}\label{eq:u_incom}
u(z,\bar z) &=& 2 \alpha \partial_z \psi + 2\partial_{\bar z} \Pi= 2 \alpha \partial_z \psi +\alpha \mathcal I(z,\bar z)\;,
\end{eqnarray}
where 
\begin{eqnarray}
\mathcal I(z,\bar z) = -\frac{2}{\pi} \int dz'd\bar z'  \frac{1}{(\bar z-\bar z')} \textrm{Re} (\partial_{z'}^2\psi)
\end{eqnarray}
arises from  the pressure gradient. The first term on the RHS of Eq.~\eqref{eq:u_incom} is the defect self-propulsion, which has the same form for both compressible and incompressible flows. Solving for the pressure field induced by an arbitrary defect configuration quickly becomes analytically intractable. The calculation can be carried out, however,  for simple configurations, such as a single defect far from all others (``isolated'' defect) and a defect dipole.  These  cases are discussed below, with technical details given in \ref{Appendix}.

\paragraph{Isolated defect.}
We consider a single positive defect located at the origin with a fixed external phase $\theta_0$. The distortions generated by the defect produce a surrounding flow determined by Eq.~\eqref{eq:u_incom}, with the induced pressure gradient calculated explicitly in~\ref{App-OneDefect}. The velocity from the active stress $u = 2\alpha\partial_z\psi$ follows from Eq.~\eqref{eq:psi}. The resulting incompressible flow field at distances $|z|$ larger than the core size is then
\begin{eqnarray}\label{eq:u_positive}
u(z,\bar z) &=&  \frac{\alpha}{\sqrt{z\bar z}}e^{2i\theta_0}+\frac{\alpha}{2}\left[ \frac{1}{\bar z} \sqrt{\frac{z}{\bar z}} e^{-2i\theta_0}-\frac{1}{\sqrt{z\bar z}}e^{2i\theta_0}\right]\;.
\end{eqnarray}

The first term on the RHS of Eq.~\eqref{eq:u_positive} is the same as obtained for compressible flow and remains finite at the defect core, contributing to the defect's propulsive velocity. The two terms in square brackets arise from pressure gradients. The second of these two terms yields a correction to the flow at the core that reduces the defect's propulsion. The  net self-propulsion velocity of the positive defects due to both active stress and pressure is then given by 
\beq
u^{+} =  2\alpha e^{2i\theta_0} +\alpha \mathcal{I} (0)=\frac{3}{4}\alpha e^{+}\;,
\eeq
where  $e^+$ is the defect polarity from Eq.~\eqref{eq:polarity+}. Pressure gradients drive defect motion in the direction opposite to that determined by active stresses, hence reducing the  the magnitude of the defect's propulsion velocity as compared to compressible flows (Eq.~\eqref{eq:v_comp}). The net propulsion remains, however, in the direction determined by the sign of activity. Finally, the shear flow due to pressure depends on derivatives of the $\mathcal{I}$-function, which vanishes at the positive defect (see \ref{App-OneDefect}), and so has no effect on an isolated positive defect. 
 
\begin{figure}[t]
\includegraphics[width=0.8\textwidth]{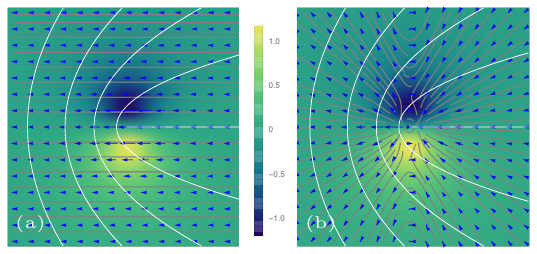}
\caption{Velocity field (gray streamlines and blue arrows) and  vorticity (color scale) induced by an isolated positive defect (Eq.~\eqref{eq:u_positive} with $\theta_0 = 0$), with the nematic texture shown as white lines for an extensile system ($\alpha= -1$) for 
(a) compressible and (b) incompressible flows.}
\label{fig:positive_velocity}
\end{figure}

The relation between the defect's self-propulsion velocity and the defect polarity remains valid for a slowly-varying orientation $\tilde\theta(z,\bar z)$. The net velocity of an isolated positive defect in an incompressible active nematic is then given by
\begin{eqnarray} 
v^+ &=&  -8 K_\lambda i \partial_{\bar z} \tilde\theta|_{z=z^+} + \frac{3}{4}\alpha e^+.
\end{eqnarray}
%

\begin{figure}[t]
\includegraphics[width=0.8\textwidth]{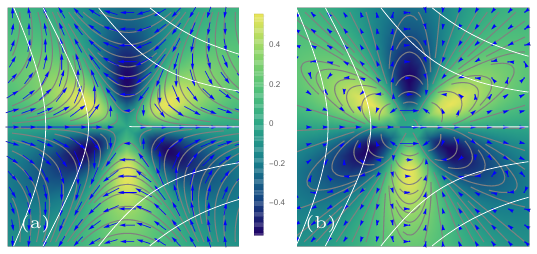}
\caption{Velocity field (gray streamlines and blue arrows) and  vorticity (color scale) induced by an isolated negative defect (Eq.~\eqref{eq:u_negative} with $\theta_0=0$), with the nematic texture shown as white lines for an extensile system ($\alpha=-1$) for 
(a) compressible and (b) incompressible flows.}
\label{fig:negative_velocity}
\end{figure}

For an isolated negative defect, the computation of the $\mathcal{I}$-function corresponding to the pressure gradient is detailed in \ref{App-OneDefect}.  The incompressible flow field generated by a negative defect outside of its core is 
\begin{eqnarray}\label{eq:u_negative}
u(z) &=&-\alpha\frac{1}{z}\sqrt{\frac{\bar{z}}{ z}} e^{2i\theta_0}-\frac{\alpha}{2}\left[\frac{z}{\bar z^2}\sqrt{\frac{z}{\bar z}}e^{-2i\theta_0} - \frac{1}{z}\sqrt{\frac{\bar z}{z}} e^{2i\theta_0} \right]\;,
\end{eqnarray}
where again the first term is the flow induced by the active stress (which vanishes at the core) and the  terms in square brackets arise from pressure gradients. 
The net velocity of an isolated $-1/2$ defect is then the same as in compressible nematic and determined solely by the local nematic distortion, %
\begin{eqnarray} 
v^- &=&  8 K_\lambda i \partial_{\bar z} \tilde\theta|_{z=z^-}.
\end{eqnarray}
 On the other hand, when other defects are present, the $\mathcal{I}$-integral is finite at the defect core, as we see next. Similar expressions for the incompressible flow field around an isolated defect have been derived  in Ref. \cite{pismen2017viscous} by solving for the Green's function in the real space.  

In Figs.~\ref{fig:positive_velocity} and~\ref{fig:negative_velocity} we compare the flow field induced by a single defect  in a compressible (panel (a)) and an incompressible (panel (b)) active nematic. In the compressible case, the flow is determined entirely by the active stress ($u = 2\alpha\partial_z\psi$). In the limit of zero viscosity  considered here, the vorticity field remains the same with or without the incompressibility constraint, since the pressure gradient is curl-free.

\subsection{Defect dipole}
\begin{figure}[t]
\includegraphics[width=0.5\textwidth]{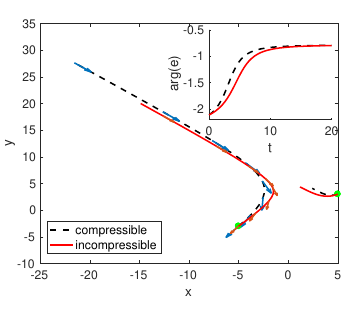}
\caption{The trajectories of a neutral defect pair sketched in Fig. \ref{fig:polarization}(c) with $\theta_0= \pi/4$ and the extensile activity $\alpha=-1$, and determined by Eqs.~\eqref{eq:dipole_incom} for an incompressible nematic (solid red lines) are compared to those obtained in a compressible fluid (dashed black lines).  The green dots are the initial position of  the two defects. The blue arrows represent the instantaneous direction of the polarity of the $+1/2$ defect.  Inset: the time evolution of the orientation of the $+1/2$ defect polarity relative to the $x$ axis for compressible (back dashed line) and incompressible (solid red line) flow.}
\label{fig:dipole_trajectory}
\end{figure}

Another configuration for which one can obtain an analytic solution is a pair of oppositely-charged defects separated  by $w = z_- -z_+$ and embedded in an otherwise uniform nematic orientation $\theta_0$ (see Fig.~\ref{fig:dipole_trajectory}(a)). The details of the evaluation of the integrals arising from pressure gradients   are presented in \ref{App-Dipole}.  The flow velocity  at the defect cores are given in terms of the relative distance $w$ as
\bea
u^+ &=& \alpha\frac{3}{2}\sqrt{\frac{\bar w}{w}} e^{2i\theta_0}-\alpha\frac{4w}{3 \bar w^2} e^{-2i\theta_0}\;,
\label{eq:uplus-dip}\\
u^- &=& -\alpha \frac{4w}{15 \bar w^2} e^{-2i\theta_0}\;.
\label{eq:uminus-dip}
\eea
The first term on the RHS of Eq.~\eqref{eq:uplus-dip} is the  self-propulsion of the $+1/2$ directed along its polarity $e^+= 2e^{2i\theta_0} \sqrt{\bar w/w}$.  Note that the polarity of the $+1/2$-defect depends on the  orientation of the negative defect through the factor $\sqrt{\bar w/w}$. This dependence survives even when the $-1/2$ defect is  infinitely far away  ($|w|\rightarrow\infty$). The second term arises from  the non-local hydrodynamic interaction between the two defects due to pressure gradients. The velocity of the $-1/2$ given in Eq.~\eqref{eq:uminus-dip} is determined entirely by hydrodynamic interactions. The different geometric structure of the two defects renders these interactions non-reciprocal, with the velocity of the negative defect  five times smaller than that of the positive one. In the limit where one defect is moved to infinity relative to the other,  the hydrodynamic interaction terms vanish and we recover zero drift for the isolated $-1/2$ defect and the self-propulsion velocity for the isolated $+ 1/2$ defect.

Substituting the explicit expressions for the incompressible flow in terms of the defect positions and polarity in  Eqs.~\eqref{eq:v_general}, we obtain 
\begin{eqnarray}\label{eq:dipole_incom}
\bm v^{+}&=& \frac{2 \bm{\hat r}}{|\bm r^{-}-\bm r^{+}|} +\frac{2\alpha}{3|\bm r^{-}-\bm r^{+}|}\left[2 \left(\bm{e^+}\cdot\bm{\hat r} \right)\bm{\hat r}-\bm{e^+}\right] + \alpha \frac{3}{4} \bm{e^+} \;,\nonumber\\
\bm v^{-}&=& -\frac{2 \bm{\hat r}}{|\bm r^{-}-\bm r^{+}|}+\frac{2\alpha}{15|\bm r^{-}-\bm r^{+}|}\left[2 \left(\bm{e^+}\cdot\bm{\hat r} \right)\bm{\hat r}-\bm{e^+} 
\right] \;,
\end{eqnarray}
where $\bm{\hat r}\equiv (\bm r^{-}-\bm r^{+})/|\bm r^{-}-\bm r^{+}|$.
The first terms on the RHS of each of Eqs.~\eqref{eq:dipole_incom} represent the familiar Coulomb interactions among the two defects. The other terms arising from corrections to active flows induced by pressure gradients are non-reciprocal in nature and yield forces normal to the line joining the two defects. 
We emphasize that these forces are distinct from non-reciprocal forces obtained in Ref.~\cite{vafa2020multi} from a perturbative analysis of multidefect interactions in a \emph{compressible} active nematic. These forces are a new result of our work. Finally, notice that there is no need to evolve the defect polarization separately, since it can be reconstructed directly from the defect positions and the uniform nematic phase $\theta_0$. 

Figure~\ref{fig:dipole_trajectory}(b) shows the  trajectories of the $\pm 1/2$ defects determined by Eqs.~\eqref{eq:dipole_incom}. For comparison we also plot the defect trajectories in a compressible flow field driven by the active stress alone. Incompressibility reduces the transient active torque  and the   velocity of the positive defect, while slightly increasing the  velocity of the negative defect. A generalization to many defects presents many challenges since the pressure field encodes all the non-local interactions between defects through integrals with many more complicated branch cuts and singularities.

\section{Defect Hydrodynamics} \label{Sec:FieldEqs}
We have derived ordinary differential equations describing the dynamics of defects as quasiparticles. In the turbulent-like nematic state where defects proliferate, it is useful to formulate a continuum model of the defect gas that describes the dynamics at length scales large compared to the mean defect separation. This was carried out in Ref. ~\cite{shankar2019hydrodynamics} for an overdamped  compressible active nematic. Here we generalize to the experimentally important case where the fluid flow is incompressible.

We describe a defect configuration in terms of microscopic density fields: defect charge and number densities, $\hat{\rho}$ and $\hat{n}$, their associated current densities, $\mathbf{\hat{J}}^{(\rho)}$ and $\mathbf{\hat{J}}^{(n)}$, and defect polarization density $\mathbf{\hat{p}}$:
\begin{eqnarray}\label{eq:micro_fields}
\hat\rho(\rb,t;\{\rb_\beta\})&=&\sum_\beta q_\beta~\delta(\rb-\rb_\beta(t))\;,\\
\hat{n}(\rb,t;\{\rb_\beta\})&=&\sum_\beta \delta(\rb-\rb_\beta(t))\;,\\
\bm{\hat{J}}^{(\rho)}(\rb,t;\{\rb_\beta\})&=&\sum_\beta q_\beta \vb_\beta~\delta(\rb-\rb_\beta(t))\;,\\
\bm{\hat{J}}^{(n)}(\rb,t;\{\rb_\beta\})&=&\sum_\beta \vb_\beta~\delta(\rb-\rb_\beta(t))\;,\\
\bm{\hat{p}}(\rb,t;\{\rb_\beta\})&=&\sum_\beta \bm{e}_\beta~\delta(\rb-\rb_\beta(t))\;. \label{eq:polarization_1}
\end{eqnarray}
where we use a hat to distinguish microscopic fields from coarse-grained ones  obtained by averaging over ensembles of defect configurations. The microscopic charge density satisfies the requisite conservation law.

The coarse-graining method consists of taking ensemble averages over different defect configurations described by a  distribution function  $P(\{\rb_\beta\})$, 
\beq
A(\rb,t) \equiv \langle \hat A(\rb,t; \{\rb_\beta\}) \rangle = \int \prod_\beta d\rb_\beta  \hat A(\rb; \{\rb_\beta\})P(\{\rb_\beta\})\;.
\eeq
and factorizing high moments to close the equations. 

To obtain the defect hydrodynamic equations, we proceed in two steps. First,  we consider a tagged defect moving in the effective mean-field induced by the rest of the defects that is described by a coarse-grained flow field $\bm{u}(\rb)=\langle\hat \ub(\rb, t;\{\rb_\beta\})\rangle$ and a coarse-grained  nematic texture $\bm v_n (\rb)=\langle\bm\nabla \hat{\tilde\theta}(\rb,t; \{\rb_\beta\})\rangle$. The dynamics of a defect in this mean field is then governed by the equations (for simplicity we consider the case with no flow alignment $\lambda=0$)
\bea \label{eq:v_coarsegrained}
&&\vb_\beta^+= 4 \bm v_n^{\perp}(\rb_\beta)+\ub(\rb_\beta) \;,\\
&&\vb_\beta^-=-4 \bm v_n^{\perp}(\rb_\beta) + \ub(\rb_\beta) \;,\\
&& \bm{\dot{e}}_\beta= - 2\bm{e}_\beta^\perp \left[\vb_\beta\cdot \bm v_n(\rb_\beta)\right]-2\bm{e}_\beta^\perp\sum_\gamma
q_\gamma\vb_\gamma\cdot\bm{f}_{\beta\gamma}^\perp\;, \label{eq:micro}
\eea
where for any vector $\bm{V}=(V_x,V_y)$ we denote the clockwise rotation corresponding to the multiplication with the Levi-Civita tensor $\epsilon_{ij}$ as $\bm{V}^\perp=\bm\epsilon\cdot\bm{V}=(V_y,-V_x)$. These equations 
couple defect dynamics to the coarse-grained flow and nematic texture, with  $\bm f_{\beta\gamma} = \rb_{\beta\gamma}/|\rb_{\beta\gamma}|^2$, which $\rb_{\beta\gamma} = \rb_\beta-\rb_\gamma$, which is the interaction force which relates the nematic distortion with defect density as 
\beq
\bm v_n^\perp(\rb_\beta)= [\bm\nabla^\perp \theta(\rb)]_{\rb=\rb_\beta}\equiv \sum_{\gamma\not=\beta} q_\gamma\bm f_{\beta\gamma}
\eeq
is the interaction force from the defect at position $\rb_\gamma$ on the defect at position $\rb_\beta$.

The phase gradient $\bm v_n$ contains the singular phase induced by all defects and relates to the macroscopic charge density $\rho$ through Stokes' theorem applied in Eq.~\eqref{eq:charge},
\bea\label{eq:vn_singular}
\bm\nabla^\perp \cdot\bm v_n  = - 2\pi \rho(\bm r)\;.
\eea
This determines $\bm v_n$ up to a potential vector field which, in the quasistatic approximation, satisfies
\beq
\bm\nabla (\bm\nabla \cdot \bm v_n  -\bm u\cdot \bm v_n+ \omega) = -2\pi \bm J^{(\rho) \perp}\;. 
\eeq
The macroscopic fluid flow $\ub$ follows the Stokes Eq.~\eqref{eq:u_stokes_unrescaled}, but with the active stress expressed in terms of the density and polarization density of positive defects,
\bea
\ub-\eta\nabla^2\ub = -\bm\nabla\Pi+\alpha\frac{\bm{p}}{n_+}\;.
\eea
Incompressibility additionally requires $\bm\nabla\cdot\ub=0$. 

The continuum equations for the defect densities and polarization density can then be obtained by taking the time derivative of the microscopic fields, Eqs.~\eqref{eq:micro_fields}, using the defect dynamics given in Eqs.~\eqref{eq:v_coarsegrained}, and applying the coarse-graining procedure. 
 Neglecting defect creation and annihilation, the equations for the defect density fields $\rho$ and $n$ take the form 
\bea
\partial_t \rho  &=& - \bm\nabla\cdot \bm J^{(\rho)}\;,
\label{eq:rho_hydro}\\
\partial_t n  &=& - \bm\nabla\cdot \bm J^{(n)}
\label{eq:n_hydro}\;.
\eea
In the presence  of defect annihilation/creation, the RHS of Eq.~\eqref{eq:n_hydro} needs to be supplemented by the kinetic rates describing these processes. The  macroscopic current densities are given by
\bea
\bm{J}^{(\rho)}(\rb,t) &=& \left\langle \sum_\beta \left[8 q_\beta \bm v_n^\perp(\rb_\beta) + \bm u(\rb_\beta) \right] q_\beta\delta(\bm r-\bm r^{(\beta)})\right\rangle \nonumber\\
&=& 2 \bm v_n^\perp n(\bm r,t) + \bm u \rho(\bm r,t)\;,\nonumber\\
\bm J^{(n)} (\rb,t) &=& \left\langle\sum_\beta  \left[8 q_\beta \bm v_n^\perp(\rb_\beta) + \bm u(\rb_\beta) \right] \delta(\bm r-\bm r^{(\beta)})\right\rangle\nonumber\\
&=& 8\bm v_n^\perp \rho (\bm r,t) + \bm u n(\bm r,t)\;.
\eea
The evolution of the microscopic polarization density follows from taking the time derivative of Eq.~\eqref{eq:polarization_1}, using Eqs.~\eqref{eq:v_coarsegrained} and expressing the sums over defects in terms of the microscopic densities and their current densities, with the result 
\begin{align}
\partial_t\hat{\bm p} +\bm U\cdot\bm\nabla \hat{\bm p} =- 2(\bm v_n\cdot\bm U)\hat{\bm p}^\perp-[\bm\nabla\cdot\bm U] \hat{\bm p}-2 \hat{\bm p}^\perp(\rb)\int d\rb' \bm{f}^\perp(\rb-\rb')\cdot \hat{\bm J}^{(\rho)}(\rb')\;.\nonumber\\
\label{eq:phat}
\end{align}
where $\bm U = \ub+4\bm v_n^\perp$ is the net flow including the one generated by nematic distortion. Upon coarse graining, we assume $\langle\hat{\bm{p}}(\rb)\hat\rho(\rb')\rangle\approx{\bm{p}}(\rb)\rho(\rb')$ and $\langle\hat{\bm{p}}(\rb)\hat{n}(\rb')\rangle\approx {\bm{p}}(\rb)n(\rb')$, and obtain the continuum equation
\begin{flalign}\label{eq:p-final}
\partial_t \bm p+\bm U\cdot\bm\nabla \bm p=
-2(\bm v_n\cdot\bm U) \bm p^\perp- [\bm\nabla \cdot \bm U]\bm p
 -2\bm p^\perp(\rb)\int d\rb' \bm{f}^\perp(\rb-\rb')\cdot \bm J^{(\rho)}(\rb')\;.\nonumber\\
\end{flalign}

The polarization equation does not contain  terms controlling growth or decay of the magnitude of $\bm{p}$  nor any stiffness/diffusive terms because we  assumed $|\mathbf{e}_\beta|=2$ (not normalized) and did not include noise in the defect dynamics.
The first term on the RHS of Eq. \eqref{eq:p-final} controls changes in the direction of polarization and arises from the active  torque in the evolution of $\mathbf{e}_\beta$.   The next term only appears when the flow is compressible and it describes the fact that density variations associated with compressible flow can provide sources and sinks of polarization. In addition, Eq. \eqref{eq:vn_singular} implies that there are also local changes in polar order depending on the charge density field, $\rho$. Finally, the last term on the RHS of Eq.~\eqref{eq:p-final} is a new nonlocal correction to the torque that arises from the  long-range nature of nematic distortions.  

It is useful to compare Eq.~\eqref{eq:p-final} for the polarization density to the one obtained in  Ref. \cite{shankar2019hydrodynamics} for a compressible fluid. As mentioned above, here we do not have terms describing polarization growth/relaxation.   The kinematic terms describing coupling of polarization to flow do reduce to those obtained in Ref.~\cite{shankar2019hydrodynamics} for compressible fluids. This can be seen by noting that in Ref.~\cite{shankar2019hydrodynamics} 
the polarization equation   was obtained by  eliminating $\ub$  from the defect dynamics  before coarse-graining, using $\ub\rightarrow \alpha{\bm{e}}_\beta$ in Eq.~\eqref{eq:phat}. If we use this in Eq. \eqref{eq:phat}, and then coarse grain, using the same moment closure as in Ref.~\cite{shankar2019hydrodynamics}, we recover the same results.  

The continuum equations obtained here  apply generally to describe the coupling of defect dynamics to any flow field. 
They resemble the equations for self-propelled polar particles in a fluid, with the difference that the $+1/2$ defects are advected and rotated not only by the fluid flow velocity $\ub$ but also by the Magnus-type force proportional to $\bm v_n^\perp$.  The fact that only positive defects act as a source of flow in the Stokes equations for $\ub$ is consistent with the fact that only the $+1/2$ defects behave like   active quasiparticles. The negative defects are advected and rotated by flows, like passive particles, but also contribute to the phase deformations described by $\bm{v}_n$.

\section{Conclusions}
\label{Sec:conclusion}
To summarize, we have analyzed the effect of arbitrary fluid flow on the motion of $\pm 1/2$ nematic defects and derived both equations describing the dynamics of defects as quasipartciles and hydrodynamic equations describing the dynamics of the defect gas on length scales large compared to the mean defect separation. By adapting the Halperin-Method formalism to a nematic order parameter, we demonstrate that it provides an effective alternative to perturbative methods of matched asymptotic expansions for deriving defect kinematics, especially for incompressible flows.
We reproduce previous results for friction-dominated, compressible flows and find new effects for incompressible systems, where pressure gradients incorporated through the incompressibility constraint yield  long-range forces among any two defect pair. We specifically examine the case of  a defect dipole  embedded in a uniform far-field texture, where analytical calculations are possible. We show that in this case the flows arising from pressure gradients reduce the self-propulsion of the $+1/2$ defects and introduce non-reciprocal forces acting on both defects. A future challenge is to extend our work to describe collective behavior and pattern formation of many interacting defects.


\ack We are thankful to Supavit Pokawanvit, Suraj Shankar, Farzan Vafa, and Zhihong You for many stimulating discussions. This work was partly supported by the National Science Foundation under Grants No. DMR-1938187 (MCM) and PHY-1748958 (LA, ZC, MJB) at the Kavli Institute for Theoretical Physics. 


\appendix 
\section{$I$-functions} \label{Appendix}
\subsection{Isolated defect:} \label{App-OneDefect}

The $I$-function corresponding to an isolated defect placed at the origin can be  evaluated at the defect position by expressing the complex area integral in polar coordinates $z' = r e^{i\theta}$. Letting  $\hat z = e^{i\theta}$, we rewrite the $I$-function as
\beq\label{eq:I_int}
\mathcal{I} (0) = \frac{2}{\pi i}\int_a^R  dr  \oint_\gamma d\hat z F(\hat z, r) ,
\eeq
where $\gamma$ is a contour around the unit circle centered at the origin. The function $F(\hat z,r)=\textrm{Re} (\partial_{z'}^2\psi)$ can be evaluated using the ansatz for $\psi$ in Eq. \eqref{eq:psi} and the fact that we are integrating from $r>a$ (distances larger than the core size) where $S(|z|)\approx 1$. For a positive defect this gives
\beq
F_+(\hat z,r) = \textrm{Re} (\partial_{z'}^2\psi_+) =-\frac{1}{8 r^2}\left(\frac{1}{\hat z} e^{2i\theta_0}+\hat z e^{-2i\theta_0}\right).
\eeq
Inserting this into Eq.~\eqref{eq:I_int} and using the residue theorem, we arrive at 
\begin{eqnarray}
\mathcal{I}(0) = -\frac{1}{4\pi i}\int_a^R  \frac{dr}{r^2}  \oint_\gamma d\hat z \left(\frac{e^{2i\theta_0}}{\hat z} +\hat ze^{-2i\theta_0} \right)\nonumber\\
= -\frac{1}{2a} e^{2i\theta_0},
\end{eqnarray}
when $R\gg a$. The $I(z,z')$-function for an isolated positive defect is computed using the transformation to polar coordinates and applying the residue theorem to the contour integral: 
\begin{eqnarray}\label{eq:I_int_pos}
\mathcal{I}_+ (z,\bar z) &=& \frac{2}{\pi i}\int_0^\infty  dr  \oint_\gamma \frac{d\hat z}{\hat z}  \frac{1}{\hat z^{-1}-\bar zr^{-1}} F_+(\hat z, r)\nonumber\\
&=&- \frac{1}{\pi i}\frac{1}{4}\int_0^\infty  \frac{dr}{r^2}  \oint_\gamma d\hat z \frac{1}{\hat z (\hat z^{-1}- \bar zr^{-1})}\left(\frac{1}{\hat z} e^{2i\theta_0}+\hat z e^{-2i\theta_0}\right)\nonumber\\
&=& -\frac{1}{2}\left[ -\frac{|z|}{\bar z^2}e^{-2i\theta_0} + \frac{1}{|z|}e^{2i\theta_0}\right].
\end{eqnarray}
The derivative of the $I_+$-function, and therefore the shear rate,  vanishes at the defect core:
\begin{eqnarray}
\partial_{\bar z}\mathcal I_+(z,\bar z)|_{z=0} &=&  \frac{2}{\pi} \int dz'd\bar z'  \frac{1}{\bar z'^2} \textrm{Re} (\partial_{z'}^2\psi_+) = \frac{2}{i\pi} \int_a^R \frac{dr}{r} \oint_\gamma d\hat z \hat z F_+(\hat z, r)\nonumber\\
&=&  -\frac{2}{i\pi} \frac{1}{8}\int_a^R \frac{dr}{r^3} \oint_\gamma d\hat z \hat z \left(\frac{e^{2i\theta_0}}{\hat z} +\hat ze^{-2i\theta_0}\right) \nonumber\\
&=& 0.
\end{eqnarray}

For a negative defect, the corresponding $F_-(\hat z,r)$ in Eq. \eqref{eq:I_int} reads 
\begin{eqnarray}
F_-(\hat z,r) = \textrm{Re} (\partial_{z'}^2\psi_-)=  \frac{3}{8 r^2} \left(\frac{1}{\hat z^3} e^{2i\theta_0}+\hat z^3 e^{-2i\theta_0}\right), 
\end{eqnarray}
and, by the residue theorem, the pressure gradient at a negative negative defect vanishes:  
\begin{eqnarray}
\mathcal{I}_- (0) &=& \frac{2}{\pi i}\frac{3}{8}\int_a^R  \frac{dr}{r^2}  \oint_\gamma d\hat z \left(\frac{1}{\hat z^3} e^{2i\theta_0}+\hat z^3 e^{-2i\theta_0}\right)\nonumber\\
&=& 0.
\end{eqnarray}
Similarly the shear flow, determined by the derivative of the $I$-function, vanishes at the defect position:
\begin{eqnarray}
\partial_{\bar z}\mathcal I_-(z,\bar z)|_{z=0} &=&  \frac{2}{i\pi} \int_a^R \frac{dr}{r} \oint d\hat z \hat z F_-(\hat z, r)\nonumber\\
&=&  \frac{1}{i\pi} \frac{3}{4}\int_a^R \frac{dr}{r^3} \oint d\hat z \hat z \left(\frac{e^{2i\theta_0}}{\hat z^3} +\hat z^3 e^{-2i\theta_0}\right) \nonumber\\
&=& 0,
\end{eqnarray}
The pressure field away from the negative defect is given by
\begin{eqnarray}\label{eq:I_int_neg}
\mathcal{I}_- (z,\bar z) &=& \frac{1}{\pi i}\frac{3}{4}\int_0^\infty  \frac{dr}{r^2}  \oint_\gamma d\hat z \frac{1}{\hat z^{-1} - \bar z r^{-1}}\left(\frac{1}{\hat z^4} e^{2i\theta_0}+\hat z^2 e^{-2i\theta_0}\right)\nonumber\\
&=& \frac{1}{2}\left[- \frac{|z|^3}{\bar z^4}e^{-2i\theta_0} + \frac{\bar z^2}{|z|^3}e^{2i\theta_0}\right].
\end{eqnarray}

\subsection{Dipole:}\label{App-Dipole}
We consider a positive defect placed at the origin, $z_+ =0$, and a negative defect at a distance $w \equiv z_- - z_+$, such that the induced nematic field is given by
\bea \label{eq:psi_dipole}
\psi_+ = \left(\frac{z}{\bar z}\right)^{\frac{1}{2}} \left(\frac{z-w}{\bar z - \bar w}\right)^{-\frac{1}{2}}e^{2i\theta_0}.
\eea
This implies that 
\begin{eqnarray}\label{re_partial2_psi}
\textrm{Re} \left(\partial_z^2 \psi_+\right) &=& -\frac{w(w-4z)}{8z^2(w-z)^2}\sqrt{\frac{z}{\bar z}}\sqrt{\frac{\bar z-\bar w}{z-w}}e^{2i\theta_0}+ c.c..
\end{eqnarray}
The area integral in Eq. \eqref{eq:I_int} for $I(0)$ can then be solved analytically using the following basic steps. First rescale the integration coordinates $z'=z/w$ and $\bar z' = \bar z /\bar w$ to simplify Eq.~\eqref{re_partial2_psi}, with lengths in units of $w$. Next make the coordinate transformation   $u^2 = \frac{z'}{z'-1}$, and $\bar u^2 = \frac{\bar z'}{\bar z'-1}$ to lift the square-root singularities to simple poles. 
The residue theorem is applied to the polar representation of the integral with the substitutions $u= r \hat u$ and $\bar u = r/\hat u$, where $\hat u$ is the complex variable associated with the contour integral over the unit circle. After these transformations the area integral becomes 
\begin{eqnarray}\label{eq:dipole_I0}
\mathcal{I}(0) &=& \frac{2}{i \pi w} \int_a^R r dr \oint_\gamma \frac{d \hat u}{\hat u} F_1(r\hat u)+\frac{2w}{i \pi \bar w^2} \int_a^R r dr \oint_\gamma \frac{d \hat u}{\hat u} F_2(r\hat u) , 
\end{eqnarray}
where 
\begin{eqnarray}
F_1(r,\hat u) &=& e^{2i\theta_0} \frac{(r^2 \hat u^2-1) (1 + 3 r^2 \hat u^2)\hat u^2}{4 r^4 (r^2-\hat u^2)}\nonumber\\
F_2(r,\hat u) &=& e^{-2i\theta_0}\frac{(r^2-\hat u^2)^2 (\hat u^2 + 3 r^2)}{4 r^4 \hat u^2 (r^2 \hat u^2-1)^2} \quad .
\end{eqnarray}
Notice that the first integral over the unit circle has poles at $\hat u = \pm r$ when $r<1$ and the second integral has poles at $\hat u = 0$ for all $r$'s and at $\hat u = \pm 1/r$ for $r>1$. We thus use the residue theorem to compute the complex integration, and then perform the integration over $r$ with the appropriate bounds set by the poles. In the limit of a small core size this yields 
\begin{eqnarray}
\mathcal{I}_+(0)=-\frac{1}{2}\sqrt{\frac{\bar w}{w}} e^{2i\theta_0}-\frac{4w}{3 \bar w^2} e^{-2i\theta_0},
\end{eqnarray}
where the first term modifies the polarity-driven self propulsion and the second term accounts for the additional drift in the  pressure field generated by the negative defect. Choosing a coordinate such that $w>0$, we see that the second term is in the direction obtained by reflecting the +1/2 defect polarity with respect to the $x$ axis, or in general, the line that connects the defects. Note that in this case, the polarity of the +1/2 defect is $-2e^{2i\theta_0}$.

In the same way to determine the pressure gradient acting on the negative defect at the origin $z_- =0$ and a distance $w \equiv z_+ -z_-$ from the positive defect, we use the order parameter $\psi$ for t  centered at the negative defect  
\bea
\psi_- = \left(\frac{z}{\bar z}\right)^{-\frac{1}{2}} \left(\frac{z-w}{\bar z - \bar w}\right)^{\frac{1}{2}}e^{2i\theta_0}
\eea
and 
\begin{eqnarray}\label{re_partial2_psi}
\textrm{Re} \left(\partial_z^2 \psi_-\right) &=& \frac{w (3 w - 4 z)}{8 (w - z)^2 z^2} \sqrt{\frac{w - z}{
 \bar w - \bar z}} \sqrt{\frac{\bar z}{z}}e^{2i\theta_0}+c.c.
\end{eqnarray}
The $I(0)$-integral from Eq.~\eqref{eq:dipole_I0} has the corresponding functions $F_1(r,\hat u)$ and $F_2(r,\hat u)$ given by
\begin{eqnarray}
F_1(r,\hat u) &=& -e^{2i\theta_0}\frac{(r^2 \hat u^2-1) (3 + r^2 \hat u^2)}{4 r^4 (r^2-\hat u^2) \hat u^2}\nonumber\\
F_2(r,\hat u) &=&- e^{-2i\theta_0}\frac{( r^2-\hat u^2)^2 (3 \hat u^2+ r^2) \hat u^2}{4 r^4 (r^2 \hat u^2-1)^2} .
\end{eqnarray}
The first function acquires poles at $\hat u=0$ for all r and at $\hat u=\pm r$ for $r<1$, while the second function has poles at $\hat u = \pm 1/r$ for $r>1$. Using the residue theorem and performing the integration over $r$ with the appropriate bounds, we find that there is only one drift term given by the pressure induced by the positive defect 
\begin{eqnarray}
\mathcal{I}_0(0)= \frac{4w}{15 \bar w^2} e^{-2i\theta_0},
\end{eqnarray}
which approaches zero in the limit where the positive defect is moved to infinity, corresponding to an isolated defect. As before the choice of a coordinate such that $w>0$ shows that this term is  along the reflection of the +1/2 polarity with respect to the $x$ axis. In this case the +1/2 defect polarity is $2e^{2i\theta_0}$.

\section*{References}
\bibliography{refs}

\end{document}